\documentclass[]{aa}
\usepackage{graphicx}
\usepackage{txfonts}

\newcommand{\be}{\begin{equation}}
\newcommand{\ee}{\end{equation}}
\newcommand{\pder}{\partial}

\begin{document}

\title{Reflection and dissipation of Alfv\'en waves in interstellar clouds}
\titlerunning{Alfv\'en waves in interstellar clouds}
\author{C.~Pinto$^{1}$, A.~Verdini$^2$, D.~Galli$^3$, M.~Velli$^4$}
\authorrunning{Pinto et al.}
\titlerunning{Alfv\'en waves in interstellar clouds}
\institute{
$^1$LUTH-Observatoire de Paris, 5 Place J. Janssen, 92190 Meudon, France
\\
$^2$Solar-Terrestrial Center of Excellence - SIDC, Royal Observatory
of Belgium, Bruxelles, Belgium
\\
$^3$INAF-Osservatorio Astrofisico di Arcetri,
Largo E. Fermi 5, I-50125 Firenze, Italy
\\
$^4$Dipartimento di Astronomia e Scienza dello Spazio,
Universit\`a di Firenze, Largo E. Fermi 5, I-50125 Firenze, Italy
\\
}
\offprints{}
%

\abstract
{Supersonic nonthermal motions in molecular clouds are often
interpreted as long-lived magnetohydrodynamic (MHD) waves. The propagation
and amplitude of these waves is affected by local physical
characteristics, most importantly the gas density and the ionization
fraction.}
%
{We study the propagation, reflection and dissipation of Alfv\'en waves
in molecular clouds deriving the behavior of observable quantities such as
the amplitudes of velocity fluctuations and the rate of energy
dissipation.}
%
{We formulated the problem in terms of Els\"asser variables for
transverse MHD waves propagating in a one-dimensional inhomogeneous
medium, including the dissipation due to collisions between ions and
neutrals and to a nonlinear turbulent cascade treated in a
phenomenological way. We considered both steady-state and time-dependent
situations and solved the equations of the problem numerically with an
iterative method and a Lax-Wendroff scheme, respectively.}
%
{Alfv\'en waves incident on overdense regions with density profiles
typical of cloud cores embedded in a diffuse gas suffer enhanced
reflection in the regions of the steepest density gradient, and strong
dissipation in the core's interior. These effects are especially
significant when the wavelength is intermediate between the critical
wavelength for propagation and the typical scale of the density
gradient. For larger wave amplitudes and/or steeper input spectra, 
the effects of the perpendicular turbulent cascade
result in a stronger energy dissipation in the regions 
immediately surrounding the dense core.}
%
{The results may help to interpret the sharp decrease of line width
observed in the environments of low-mass cloud cores in several
molecular transitions.}


\maketitle

\section{Introduction}
\label{intro}

The supersonic nonthermal motions observed in molecular clouds have
been attributed to the presence of magnetohydrodynamic (MHD) waves of
sub-Alfv\'enic amplitudes (Arons \& Max~1975), of which the Alfv\'en
waves are the longest-lived (Zweibel \& Josafatsson~1983).  Indeed,
molecular clouds and the cores embedded in them are threaded by the
Galactic magnetic field, as revealed by polarization maps of the
thermal dust emission (see e.g. Tang et al.~2009).

Molecular cloud cores, the sites of star formation, are characterized
by velocity dispersions with nonthermal motions subsonic and uniform,
in contrast to the surrounding lower density gas, where nonthermal
motions are supersonic and in general increasing with size (Barranco \&
 Goodman~1998; Goodman et al.~1998; Caselli et al.~2002). The NH$_3$
observations by Pineda et al.~(2010) of the B5 region in Perseus
revealed that the transition in velocity dispersion between the
quiescent core and its more ``turbulent'' environment is sharp, taking
place in a thin layer with a width of less than $0.04$~pc.

Several groups have performed simulations and developed
(semi)analytical models of the propagation of non-linear and linear
Alfv\'en waves in interstellar clouds, but the main focus of these
studies was either on the generation of density fluctuations and
velocity dispersions with properties comparable to those of molecular
clouds (Heitsch \& Burkert~2002; Kudoh \& Basu~2003, ~2006), or on the
role of internally generated Alfv\'en waves to support clumps within
molecular clouds against their own self-gravity, taking into account the
ionization structure and the ion-neutral drift (Martin et al.~1997;
Coker et al.~2000).  Our main objective is to study how density gradients, 
ionization fraction and non-linear wave interaction affect
the propagation and the amplitude of Alfv\'en waves at the small scale
of the interface between a dense core and its environment.

For this goal, we considered a simple one-dimensional cloud model for
which the propagation of Alfv\'en waves is formulated in terms of
Els\"asser variables.  This formulation allows one to treat propagation,
reflection, damping, and nonlinear interactions in a simpler form
compared to the original fluid equations (Els\"asser~1950; Dobrowolny
et al.~1980; Marsch \& Mangeney~1987).  Our approach is similar to that
adopted for studying the propagation and reflection of Alv\'en waves in
radially stratified stellar atmospheres and winds (Velli~1993).  
In this context the density gradients have the main effect of
modulating the amplitude of Alfv\'en waves. This may affect the damping
that results from the ion-neutral drift and at the same time trigger the
development of a (perpendicular) turbulent cascade. Here we will focus
mostly on the first aspect, although we will also use a phenomenological model
to describe the dissipation of turbulence (successfully applied to 
the solar wind, e.g. Verdini et al. 2010) to see how it affects the results.

The paper is organized as follows: in sect.~\ref{disp} we derive the
wave equation and review the conditions for wave propagation; in
sect.~\ref{elss} we formulate the problem in terms of Elss\"asser
variables; in sect.~\ref{code} we describe the method of solution;
we discuss some special cases in sect.~\ref{special} and, in
sect.~\ref{cloud}, we consider wave propagation, damping and reflection
in a one-dimensional cloud model with a specific density and ionization
profile; we evaluate the energy dissipation due to both ion-neutral
drift and nonlinear effects; finally in sect.~\ref{concl} we summarize
our conclusions.

\section{Wave equation and dispersion relation}
\label{disp}

We considered a one-dimensional density distribution $\rho(x)$, threaded
by a magnetic field ${\bf B}=B\hat{\bf x}$, in which the field is
stationary and uniform, the latter condition resulting from the
requirement of zero divergence, $\nabla\cdot{\bf B}=0$.  The total
density of the gas $\rho$ is the sum of the densities of neutrals
($\rho_n$), ions ($\rho_i$) and electrons ($\rho_e$), and we assumed 
that the gas is electrically neutral. For applications to molecular
clouds, we assumed that the field is frozen in the ions, which 
transfer momentum to the neutrals via elastic collisions. We neglected 
the electron's inertia and focused on the ambipolar diffusion regime
(see e.g. Pinto et al.~2008), ignoring Hall and Ohmic diffusion.  
In the regime of interest in this work ($n({\rm H}_2)\sim
10^4$--$10^5$~cm$^{-3}$, $B\sim 10$--$100$~$\mu$G), Ohmic diffusion is
about 10 orders of magnitudes smaller than ambipolar diffusion and can
be safely neglected; on the other hand, Hall diffusion is about one
order of magnitude smaller than ambipolar diffusion, for standard grain
size distribution, and about 4 orders of magnitudes smaller in the case
of large (micron size) dust grains (see, e.g., Wardle \& Ng~1999;
Nakano et al.~2002; Pinto et. al.~2008).  The consequences of the Hall
effect on the propagation of Alfv\'en wawes in non-homogeneous media
have been considered by Campos \& Isaeva~(1993).

In our analysis, we also assumed that the magnetic force is
transferred to the neutrals only by collisions with ions. Actually, in
the density regime of interest, charged dust grains can contribute as
much as ions in the frictional force, provided that a population of
small ($\lesssim \mu$m size) grains is present (Nakano et al.~2002).
If only large ($\gtrsim \mu$m size) grains are present, as expected if
some amount of dust coagulation has taken place (Ormel et al.~2009),
 the contribution of grains to the frictional force becomes much
smaller. On the other hand, large-size dust grains may induce
dispersive effects on the propagation of nonlinear magnetic
disturbances, and even promote the formation of soliton structures,
although on scales at best one order of magnitude smaller than the
typical size of a molecular cloud core (Pandey et al.~2008).

We considered transverse disturbances that are characterized by magnetic
field ${\bf b}$ and velocities ${\bf u}_n$ and ${\bf u}_i$ (the latter
were also assumed to be incompressible).  We assumed that the sources of
the waves are located a sufficiently large distance from the core's
midplane to neglect their density-pressure feedback on the cloud.  The
equations of motion and the induction equation read
\be
\rho_n\frac{\partial{\bf u}_n}{\partial t} 
-\gamma_{in}\rho_i\rho_n({\bf u}_i-{\bf u}_n) 
= - \rho_n {\bf u}_n\cdot\nabla_\bot{\bf u}_n,
\label{eqn}
\ee
\begin{eqnarray}
\lefteqn{\rho_i\frac{\partial{\bf u}_i}{\partial t} 
-\frac{B}{4\pi}\frac{\partial {\bf b}}{\partial x}
-\gamma_{in}\rho_i\rho_n({\bf u}_n-{\bf u}_i)= 
- \rho_i {\bf u}_i\cdot\nabla_\bot{\bf u}_i
-\frac{1}{8\pi}\nabla_\bot |{\bf b}|^2}
\nonumber \\
& & 
+\frac{1}{4\pi} {\bf b}\cdot\nabla_\bot{\bf b}, 
\label{eqi}
\end{eqnarray}
\be
\frac{\partial{\bf b}}{\partial t} 
-B\frac{\partial {\bf u}_i}{\partial x}
= {\bf b}\cdot\nabla_\bot{\bf u}_i - {\bf u}_i\cdot\nabla_\bot{\bf b},
\label{eqb}
\ee
where 
\be
\gamma_{in}=\frac{\langle\sigma w\rangle_{in}}{m_i+m_n},
\ee
is the collisional drag coefficient, $\langle\sigma w\rangle_{in}$ is
the momentum transfer rate coefficient, and $\nabla_\bot$ is the
gradient along the direction perpendicular to the large-scale field
$B$. On the right-hand side we have grouped the nonlinear terms: note
that because of the assumption of transverse fluctuations they involve
only gradients in that direction.

Combining the linearized
Eqs.~(\ref{eqn}), (\ref{eqi}) and (\ref{eqb}), we obtain the wave
equation for ${\bf u}_i$,
\be
\frac{\pder^3 {\bf u}_i}{\pder t^3}-V_{Ai}^2\frac{\pder^3 {\bf u}_i}{\pder x^2\pder t}
+\gamma_{in}\rho_n\left(\frac{\rho}{\rho_n}\frac{\pder^2 {\bf u}_i}{\pder t^2}
-V_{An}^2\frac{\pder^2 {\bf u}_i}{\pder x^2}\right)=0,
\label{all1}
\ee
where $\rho=\rho_n+\rho_i$ and $V_{Ai}$, $V_{An}$ are the Alfv\'en
velocities in the ions and the neutrals, respectively:
\be
V_{Ai}=\frac{B}{\sqrt{4\pi\rho_i}}, \qquad V_{An}=\frac{B}{\sqrt{4\pi\rho_n}}.
\ee
Substituting in Eq.~(\ref{all1}) a solution in the form
${\bf u}_i(x,t)={\bf u}_{i0}\exp[i(kx-\omega t)]$, we obtain the 
well-
known dispersion relation (DR)
\be
\omega[\omega^2-(k V_{Ai})^2]+i\gamma_{in}[\rho\omega^2-\rho_i(k V_{Ai})^2]=0,
\label{dr0}
\ee
first derived by Piddington~(1956) and later by Kulsrud \& Pearce~(1969).  

The amplitudes of the fluctuations in ${\bf u}_n$ and ${\bf u}_i$ are related 
to the amplitude of the fluctuations of ${\bf b}$ by
\be
{\bf u}_{0n}=-V_{Ai}\frac{i\gamma_{in}\rho_i}
{\omega+i\gamma_{in}\rho}
\left(\frac{kV_{Ai}}{\omega}\right)\frac{{\bf b}_0}{B},
\label{amp_n}
\ee
and
\be
{\bf u}_{0i}=-V_{Ai}\frac{\omega [\omega+i\gamma_{in}\rho]
-\gamma_{in}^2\rho_i\rho_n}{[\omega+i\gamma_{in}\rho]
(\omega+i\gamma_{in}\rho_n)}
\left(\frac{kV_{Ai}}{\omega}\right)\frac{{\bf b}_0}{B}.
\label{amp_i}
\ee

For $\rho_i/\rho_n \ll 1$, there are two
separated regimes of wave propagation: long waves (for $k
V_{Ai}\ll\sqrt{\rho_i/\rho_n}\gamma_{in}\rho_n$) and short waves (for
$k V_{Ai}\gg\gamma_{in}\rho_n$). The physical reason is that when the
timescale for wave damping becomes shorter than the wave period, wave
oscillation cannot occur.
Short waves are high-frequency waves propagating at the ion's Alfv\'en 
speed, damped by collisions with neutrals. 
In the short-wave regime, the ratio of the amplitudes of the velocity 
oscillations reduces to
\be
\frac{|{\bf u}_{0n}|}{|{\bf u}_{0i}|}=\frac{\gamma_{in}\rho_i}{\omega},
\ee
i.e. the neutrals are weakly coupled to the ions and their amplitude of
oscillation is very small. Therefore, these waves cannot be responsible
for the linewidths of neutral species observed in molecular clouds.

In the limit of long waves, the wave equation (\ref{all1}) becomes
\be
\frac{\pder^2 {\bf u}_i}{\pder t^2}-V_A^2\frac{\pder^2 {\bf u}_i}{\pder x^2}
-\eta_{\rm AD}\frac{\pder^3 {\bf u}_i}{\pder x^2\pder t}=0,
\label{alw}
\ee
where $V_A$ is the Alfv\'en speed in the gas of neutrals and ions 
(of density $\rho=\rho_n+\rho_i$), and
$\eta_{\rm AD}$ is the ambipolar diffusion resistivity
\be
\eta_{\rm AD}=\frac{B^2}{4\pi\gamma_{in}\rho\rho_i}.
\ee
In this regime, the DR of Eq.~(\ref{dr0}) reduces to
\be
\omega^2-k^2 V_A^2+i\eta_{\rm AD}k^2\omega=0.
\label{DR}
\ee
For the propagation and damping of Alfv\'en waves in a molecular cloud
a description in terms of {\em forced waves} is appropriate. Indeed,
there are several possible sources of hydromagnetic waves in a
molecular cloud, such as gravitational collapse of smaller parts,
large-scale non-radial oscillations of the cloud, radiation pressure from
young stars, expansion of H{\small II} regions, stellar outflows (Arons
\& Max~1975).  In this case, the DR should be considered as a quadratic
equation for the wavenumber $k$ (with real and imaginary part) as a 
function of the frequency $\omega$ (real).  Defining a complex wave
number $k=\Re(k) + i\Im(k)$, 
in the limit $\omega\ll V_A^2/\eta_{\rm AD}$, 
we derive from the real and imaginary parts
of the dispersion relation Eq.~(\ref{DR}) the expressions
\be
\Re(k)\approx \pm \frac{\omega}{V_A},\qquad\mbox{and}\qquad
\Im(k)\approx \pm \frac{\omega^2}{2 V_A^3}\eta_{\rm AD}.
\label{DRsol_appr}
\ee
The condition for wave propagation, $|\Re(k)| \ge |\Im(k)|$, is satisfied 
only if the frequency of forcing is lower than a critical
frequency $\omega_{\rm cr}=2V^2_A/\eta_{\rm AD}$, or, the wavelength is
longer than a critical value $\lambda_{\rm cr}=\pi \eta_{\rm AD}/V_A$
(Kulsrud \& Pearce~1969).
In the low-ionization regime typical of a 
molecular cloud, $\rho\approx \rho_n$, the condition for wave propagation
reduces to $\omega\le 2\gamma_{in}\rho_i$ or, in terms of wavelength, to
\be
\lambda\ge \lambda_{\rm cr}=\frac{\pi^{1/2}B}{2\gamma_{in}\rho^{1/2}_n\rho_i}=
\frac{\pi V_A}{\gamma_{in}\rho_i}.
\label{longwave}
\ee
We selected the case of depleted clouds where all metal species are frozen onto 
grains and the dominant ions are H$^+$ and H$^+_3$.
In this case $\gamma_{in}=2.38\times
10^{14}$~g$^{-1}$~cm$^3$~s$^{-1}$ (see Pinto \& Galli 2008) and the
mean mass of neutrals and ions are $m_n=2.33 m_H$ and $m_i=3m_H$,
respectively. Inserting numerical values, we obtain
\be
\lambda_{\rm cr}=3.8\times 10^{-3}~\left(\frac{B}{100~\mbox{$\mu$G}}\right)
\left(\frac{n}{10^5~\mbox{cm$^{-3}$}}\right)^{-3/2}
\left(\frac{x_i}{10^{-7}}\right)^{-1}~\mbox{pc},
\label{longwave_ext}
\ee
where $x_i=n_i/n$ is the ionization fraction. 
In the regime of long waves, the amplitudes of the velocity oscillations reduce to
\be
{\bf u}_{0n}=-\frac{k V_A^2}{\omega}\frac{{\bf b}_0}{B},
\qquad\mbox{and}\qquad 
{\bf u}_{0i}=-\frac{\omega}{k}\frac{{\bf b}_0}{B}.
\ee
Since $|k|=\omega/V_A$ to lowest order in $(\eta_{\rm
AD}\omega/V_A^2)^2$, the two velocity amplitudes are equal in modulus
$|{\bf u}_{0n}|=|{\bf u}_{0i}|=V_A|{\bf b}_0|/B$.  These
long-wavelength Alfv\'en waves propagate with speed $V_A$ in a
well-coupled ion-neutral gas, and may be responsible for the observed
linewidths of ionic and neutral species observed in molecular clouds.

\section{Model equations}
\label{elss}
The condition for the propagation of long waves is equivalent to
assume that neutrals and ions are well coupled ($|{\bf u}_i-{\bf u}_n|/V_{An}
\ll |{\bf b}|/B$).  With this condition, and the approximation 
$\rho_i \ll \rho_n$, the ion velocity can be obtained from Eq.~(\ref{eqi}) as
\be
{\bf u}_i={\bf u}_n+\frac{\eta_{\rm AD}}{B}\frac{\partial {\bf b}}{\partial
x}+\frac{\eta_{\rm AD}}{B^2}{\bf b}\cdot \nabla_\bot{\bf b}
-\frac{\eta_{\rm AD}}{2B^2}\nabla_\bot|{\bf b}|^2.
\label{equiun}
\ee
Upon substitution of the above equation into the induction
Eq.~(\ref{eqb}) one obtains a quite complicated equation. 
To make it more tractable, still accounting for non-linear interactions,
we adopted the following additional assumptions:
i) fluctuations have small amplitudes ($b/B\approx u_n/V_a<1$), thus, 
when substituting Eq.~(\ref{equiun}) into Eq.~(\ref{eqb}) we expand the 
non-linear terms only to the second order (cubic terms ${\cal O}(b^3)$ are neglected); 
ii) the ambipolar diffusion coefficient varies on a 
scale much larger than all the other quantities do, thus we neglect
terms involving its derivatives.
Under these assumptions, the induction equation is of  the form
\begin{eqnarray}
\lefteqn{\frac{\pder {\bf b}}{\pder t}=B\frac{\partial {\bf u}_n}{\partial x}
+{\bf b}\cdot\nabla_\bot{\bf u}_n -{\bf u}_n\cdot\nabla_\bot{\bf b}}
\\\nonumber
& &
+\eta_{AD}\left[
\frac{\pder }{\pder x}\left(\frac{\pder {\bf b}}{\pder x}\right)
+\frac{2}{B}{\bf b}\cdot\nabla_\bot\left(\frac{\pder{\bf b}}{\pder x}\right)
-\frac{1}{2B}\frac{\pder\nabla_\bot |{\bf b}|^2 }{\pder x}
\right].
\label{balmost}
\end{eqnarray}
We see that nonlinearities also appear in terms that contain the ambipolar
diffusion coefficient (the last two terms), therefore we expect that the
turbulent cascade is also affected by ion-neutral drift. In view of the simple
phenomenology that we will use (Eq.~(\ref{phen})) to describe the dissipation 
of turbulence, we additionally simplified the equation by adopting an ordering 
of the parallel and perpendicular length scales, namely iii)
$\lambda_{cr}/\ell_{||}<<b/B<<\ell_\bot/\ell_{||}$, where $\ell_{||,~\bot}$ 
are the typical sizes of parallel and perpendicular large scales. 
We thus obtain the final model equations:
\be
\rho_n\frac{\partial {\bf u}_n}{\partial t}=
\frac{B}{4\pi}\frac{\partial {\bf b}}{\partial x}
-\frac{1}{8\pi} \nabla_\bot|{\bf b}|^2
+\frac{1}{4\pi} {\bf b}\cdot \nabla_\bot{\bf b}
- \rho_n {\bf u}_n\cdot\nabla_\bot{\bf u}_n,
\label{uperpunid}
\ee
and
\be
\frac{\pder {\bf b}}{\pder t}=B\frac{\partial {\bf u}_n}{\partial x}+
\eta_{\rm AD}\frac{\partial}{\partial x}
\left(\frac{\partial {\bf b}}{\partial x}\right)
+{\bf b}\cdot\nabla_\bot{\bf u}_n -{\bf u}_n\cdot\nabla_\bot{\bf b}.
\label{bperpunid}
\ee

Eq.~(\ref{uperpunid}) and (\ref{bperpunid}) can be
conveniently expressed in terms of the Els\"asser variables,
\be
\label{defz}
{\bf z}_{\pm}={\bf u}_n\mp\frac{{\bf b}}{\sqrt{4\pi \rho_n}},
\ee
by dividing Eq.~(\ref{bperpunid}) by $\sqrt{4\pi \rho_n}$, then adding
and subtracting Eq.~(\ref{uperpunid}) and (\ref{bperpunid}):
\begin{eqnarray}
\lefteqn{\frac{\partial {\bf z}_\pm}{\partial t}\pm V_A
\frac{\partial{\bf z}_{\pm}}{\partial x}= 
-{\bf z}_\mp\cdot\nabla_\bot{\bf z}_\pm-\frac{1}{\rho_n}\nabla_\bot P_m
\pm\frac{V_A^\prime}{2}({\bf z}_\pm-{\bf z}_\mp)} 
\nonumber \\
& & 
\pm\frac{\eta_{\rm AD}}{2}({\bf z}_\pm-{\bf z}_\mp) 
\left[2\left(\frac{V^\prime_A}{V_A}\right)^2-
\frac{V^{\prime\prime}_A}{V_A}\right]
-\eta_{\rm AD}\frac{V_A^\prime}{V_A}
\left(\frac{\partial{\bf z}_{\pm}}{\partial x}
-\frac{\partial{\bf z}_{\mp}}{\partial x}\right)
\nonumber \\
& & 
+\frac{\eta_{\rm AD}}{2}\left(\frac{\partial^2{\bf z}_{\pm}}{\partial x^2} 
-\frac{\partial^2{\bf z}_{\mp}}{\partial x^2}\right).
\label{evolzspatialzmp} 
\end{eqnarray}
The superscripts $^\prime$ and $^{\prime\prime}$ indicate first- and
second-order spatial derivatives along the $x$ direction and $P_m$ is
the magnetic pressure.  The first two terms on the left-hand side
represent the temporal and spatial wave propagation, respectively . On
the right-hand side, the first two term account for non-linear
interactions; the term proportional to $V_A^\prime$ describes the
losses related to the presence of gradients in the Alfv\'en velocity
and wave reflection; the terms proportional to $\eta_{\rm AD}$ are
associated to the dissipation by ambipolar diffusion and combined
effects of density gradients and ion-neutral drift.  Because we assumed
fluctuations to lie in the plane perpendicular to $B$, the gradient
$\nabla_\bot$ involves only perpendicular scales.  Given the high
Reynolds numbers found in molecular clouds ($R_m\approx 10^8$, Myers \&
Khersonsky 1995), a perpendicular turbulent cascade is expected to
develop. These non-linear couplings involve only counter-propagating
waves. Note however that even if only one mode is present, say $z_+$,
density gradients or ion-neutral drift ensure the presence of the other
mode ($z_-$) and hence the triggering of the cascade.

The hypotheses that we have adopted in deriving
Eqs.~(\ref{uperpunid},~\ref{bperpunid},~\ref{evolzspatialzmp}) are summarized in
the following ordering:
\be
\frac{\lambda_{cr}}{\ell_{||}}<<\frac{b}{B}<<\frac{\ell_\bot}{\ell_{||}}<<\frac{\rho_n}{\rho_i}.
\label{hp}
\ee
Note that the first inequality is a stronger constraint than the long wavelength
condition Eq.~(\ref{longwave}), but in molecular clouds one typically finds that
$\lambda_{cr}/\ell_{||}\approx\eta_{AD}/V_AL~\lesssim0.06$ (see
Eq.~(\ref{longwave_ext})) and
$\rho_n/\rho_i\gtrsim 10^7$, therefore the above ordering is generally
satisfied even for $\ell_\bot/\ell_{||}>1$. 
The condition $b/B<<\ell_\bot/\ell_{||}$ can be rewritten as
$\tau_A/\tau_{NL}<1$ where the Alfv\'en and non-linear
timescales are given by $\tau_A=\ell_{||}/V_A$ and $\tau_{NL}=\ell_\bot/z_\pm$ 
(we assumed $b/B\approx u/V_A\approx z_\pm/V_a$). In other words, the equations
 are suited to describe \emph{weak} turbulence. 
Our inequality is more stringent than $\tau_A/\tau_{NL}\ll B/b$ given
in Galtier~et~al.~(2002), which was used to build a cascade model. We will not
attempt such a detailed description of the cascade for two reasons that are related
to the inhomogeneity.
It will be clear in the following that the density gradients couple
counter-propagating waves. A first consequence is that the phase of
oppositely propagating low-frequency waves cannot be considered as random
anymore, preventing us from exploiting a phase averaging method at large 
scales to obtain the kinetic equations.
Moreover, in the inhomogeneous case typical of the cloud-core systems, 
the density gradients act as semi-reflecting boundaries, which enhance the amplitudes, 
and therefore the cascade rate of large-scale modes.
These low-frequency waves in which energy is
accumulated pose a problem with the continuity of the low-frequency tail of the spectrum.
Our nonlinear model is really only in the spirit of von Karman and Howarth (1938), (see also Mac Low 1999), 
where the large eddies control the dissipation rate and the specific mechanism for small-scale 
dissipation is not of central importance. Large-scale fluctuations control the cascade, and fast, 
small-scale dissipation mechanisms, whether collisional or not, 
act as passive absorbers of cascaded energy. \\
Another comment concerns the possibility of nonlinear compressible
 interactions and processes such as parametric decay.
These interactions may play a role, but in a completely turbulent situations
where fluctuations have completely random
phases, the decay via these resonant processes is greatly reduced (Malara \& Velli 1996). 
Consequently, the density fluctuations that are there will play a role via processes such as 
phase mixing/resonant absorption, and shock heating, but will not
alter, except at shocks, the dominant nature of the energy cascade
 given by the dimensional model described below.

\section{Numerical method}
\label{code}
We first sought solutions in form of monochromatic waves for the linearized
Eqs.~(\ref{evolzspatialzmp}).
This allowed us to show
the role of interference between waves as
produced by reflection and ambipolar diffusion (see
section~\ref{special}). The numerical method is presented in the next
section~\ref{timein-scheme}.
Then we solved the full set of Eqs.~(\ref{evolzspatialzmp}) in their
time-dependent form, treating the non-linear interactions in a
phenomenological way and considering fluctuations that are characterized by a
steep frequency spectrum. The details of the
numerical method and the phenomenology adopted are given in
section~\ref{time-scheme}. 

\subsection{Time-independent scheme for linear equations}
\label{timein-scheme}
We neglected non-linear terms and hence the perpendicular structure of
Alfv\'en waves. We sought solutions for monochromatic waves of the
form ${\bf z}_\pm(x,t)=V_{A0}{\bf y}_{\pm}(x)\exp(-i\omega t)$.  For a
cloud of typical size $L$ and density $\rho_0$, we can define a
characteristic Alfv\'en speed $V_{A0}=B/\sqrt{4\pi\rho_0}$ and Alfv\'en time
$t_A=L/V_{A0}$, and the
nondimensional variables $\xi=x/L$ and $\tilde{\rho}=\rho/\rho_0$.
In nondimensional form, eqs.~(\ref{evolzspatialzmp}) can be written as
\begin{eqnarray}
\label{evol}
\lefteqn{\pm\frac{1}{\tilde\rho^{1/2}}y_{\pm}^\prime=
iWy_\pm\mp\frac{1}{4}\left(y_\pm-y_\mp\right)
\left\{\frac{\tilde{\rho}^\prime}{\tilde{\rho}^{3/2}}\pm\tilde{\eta}_{\rm AD}
\left[\frac{\tilde{\rho}^{\prime\prime}}{\tilde{\rho}}
-\frac{1}{2}\left(\frac{\tilde{\rho}^\prime}{\tilde{\rho}}\right)^2\right]\right\}}
\nonumber \\
& & 
+\tilde{\eta}_{\rm AD}\frac{\tilde{\rho}^\prime}{2\tilde{\rho}}(y_\pm^\prime-y_{\mp}^\prime)
+\frac{\tilde{\eta}_{\rm AD}}{2}(y_{\pm}^{\prime\prime}-y_{\mp}^{\prime\prime}),
\end{eqnarray}
where $W=\omega L/V_{A0}$ and $\tilde{\eta}_{\rm AD}=\eta_{\rm
AD}/(V_{A0}L)$. 
For uniform density ($\tilde\rho=1$) and without dissipation
($\tilde\eta_{\rm AD}=0$), Eq.~(\ref{evol}) describes the propagation
of Alfv\'en waves traveling to the right ($+$ sign) and to the left ($-$
sign).

Because ambipolar diffusion is a small effect in typical ISM conditions
($\tilde\eta_{\rm AD}\ll 1$), we adopted a perturbational approach. We
initially solved Eqs.~(\ref{evol}) guessing at a functional form for the
second-order derivatives. In particular, we assumed the exponentially
damped behavior predicted by analytical studies (we have verified that
the numerical results are independent of the choice of the initial
guess); then we used the second-order derivatives of the solutions
obtained by the guess function to integrate Eqs.~(\ref{evol}) and
obtained a new set of second-order derivatives. This procedure was
iterated until convergence of the solution was attained.

Special care has to be taken when choosing boundary conditions suitable 
to describe waves incident on both sides of an interstellar cloud because 
of the interference between the incoming waves.
 When only ambipolar diffusion is present, the iterative
procedure reduces the problem to a boundary-value problem for the
fields $z^\pm$.  For wave propagation in a density gradient
without dissipation, the boundary-value problem is well defined only
when one wave vanishes at one boundary.  In the more realistic case,
where both modes are injected at the boundaries, solutions depend on
the relative phase between the injected $z_\pm$. The phase in fact
controls the interference between the incident wave and the reflected
wave inside the density profile.  We will describe in detail the
appropriate initial conditions in the application to interstellar clouds (sect.~\ref{cloud}), 
first dealing with density gradients, then considering both ambipolar diffusion and density gradients.

\subsection{Time-dependent scheme for non-linear equations}
\label{time-scheme}
To evaluate the effect of turbulent dissipation, which can be triggered by non-linear wave interactions, 
we did not model the turbulent cascade. Instead 
we used a phenomenological expression (Dmitruk~et~al.~2001) that
accounts for the damping of the waves and preserves the structure of
non-linear terms, allowing interactions only among counter-propagating waves. 
We therefore integrated eqs.~(\ref{evolzspatialzmp}) modified with the following substitution in the RHS,
\be
-{\bf z}_\mp\cdot\nabla_\bot{\bf z}_\pm-\frac{1}{\rho_n}\nabla_\bot P_m 
\rightarrow -\frac{|{\bf z}_\mp|{\bf z}_\pm}{2\ell_\bot},
\label{phen}
\ee
where $\ell_\bot$ represents the dimension of the
largest eddies (the energy containing scale in the turbulent field).

The form of the nonlinear term may be heuristically derived in the framework
of reduced MHD in which 
variations along
the perpendicular directions are decoupled from those along the
magnetic field ($\nabla=\nabla_\bot+\nabla_{||}$, with
$\nabla_\bot\gg\nabla_{||}$).
In presence of a strong magnetic field, 
the propagation time of the Alfv\'en waves
$\tau_a$ is equal to or shorter than, the characteristic time-scale
 for nonlinear interaction $\tau_{NL}$ over most
of the Fourier space (consistent with the ordering in Eq.~(\ref{hp})).
The nonlinear cascade becomes anisotropic, developing preferentially
in the plane perpendicular to the direction of the mean field
(Shebalin~et~al~1983; Ougthon~et~al.~1994; Goldreich~\&~Sridhar~1995).
One can therefore Fourier-decompose Eq.~(\ref{evolzspatialzmp}) only in 
${\bf k}_\bot$ (wavevector lying in plane perpendicular to the large-scale
magnetic field). The global effect of this perpendicular
cascade can be described by means of two quantities at the large scales, 
namely a common similarity (correlation) scale and the average velocity 
difference among points belonging to the same eddy.
Identifying these two quantities with the integral turbulent length ($\ell_\bot$) and the
fluctuations' amplitude of the Els\"{a}sser fields, we can construct a
characteristic timescale $\tau^\pm_{NL}=\ell_\bot/\vec{z}_\pm$, which
accounts for nonlinear turbulent interactions in
Eq.(\ref{evolzspatialzmp}). This phenomenology assumes that all the
interactions are resonant, in practice overestimating the heating rate for high-
frequency fluctuations, which instead would have a reduced nonlinear coupling because 
of their fast decorrelation. On the other hand, the phenomenology still accounts for the reduction
of the cascade and associated damping due to the imbalance between the $\bf{z}_\pm$
fields.

In our approximation, parallel scales are not generated by the cascade
itself, as is expected to happen for strong turbulence (Goldreich~\&~Sridhar~1995). 
Therefore, here the perpendicular turbulent cascade represents a channel for
energy dissipation that is alternative to ambipolar diffusion (assumed
to act only on the parallel scale).
We finally remark that the ordering in Eq.~(\ref{hp}) is assumed to be valid at large
scales but it breaks at smaller scales, where the cascade is expected to become
strong, eventually reaching the dissipative scale. 
However, for the purpose of estimating the turbulent heating rate,
the phenomenology is still valid and consistent with the ordering 
since the overall cascade rate will be controlled by the weaker cascade at
the modeled large scales.

We integrated the evolution Eqs.~(\ref{evolzspatialzmp}), with the
substitution of Eq.(\ref{phen}), using a numerical scheme of first
order in time and second order in space (a centered Lax-Wendroff
scheme).  Passing to time-dependent equations allows us to better model
the interference (dependence on the phase) of Alfv\'en waves (from incoherent sources) 
that are assumed to enter from both sides of the cloud with random phases.  Boundary conditions are
transparent, no reflection occurs there. Waves are injected into the
cloud according to
\be
{\partial{z_+}}/{\partial t}|_{x=-L}=g(t),
\qquad
{\partial{z_-}}/{\partial t}|_{x=L}=f(t),
\ee
where the forcing $f(t)$, $g(t)$) has a spectrum peaked at low
frequencies and declining as $\omega^{-2}$. The escaping waves at the
boundaries, $\partial{z_+}/\partial t$ at $x=L$ and
$\partial{z_-}/\partial t$ at $x=-L$, are not specified and obtained as
part of the solution.

\section{Special cases}
\label{special}
To gain insight into the nature of the solution, we first considered 
the effects of ambipolar diffusion for waves propagating in a uniform
medium and then the propagation of waves in a non-uniform medium
without ambipolar diffusion.
\subsection{Damping by ambipolar diffusion, uniform density}
\label{SectAD}
For a uniform density medium, $\tilde\rho=1$, 
Eq.~(\ref{evol}) reduces to
\be
\label{evolzmad}
\pm y_{\pm}^\prime=iW y_{\pm}
+\frac{1}{2}\tilde{\eta}_{\rm AD}
(y_{\pm}^{\prime\prime}-y_{\mp}^{\prime\prime}).
\ee
In the limit of small dissipation, 
setting $\epsilon\equiv \tilde{\eta}_{\rm AD}W/4 \ll 1$,
Eq.~(\ref{evolzmad}) has a general solution of the form 
\be
y_{\pm}=(a_{\pm}-i\epsilon a_{\mp})e^{\pm iW\xi}e^{\mp \xi/x_d}
+i\epsilon a_{\mp}e^{\mp iW\xi}e^{\pm \xi/x_d},
\label{etasol}
\ee
where $x_d=(2\epsilon W)^{-1}$ is the damping length and and $a_{\pm}$
are constants. In dimensional units,
\be
\ell_d=x_d L=\frac{2V_{A0}^3}{\eta_{\rm AD} \omega^2}.
\label{damp}
\ee
Notice that the damping length can also be expressed as 
$\ell_d=2\pi\lambda^2/\lambda_{\rm cr}$.

In this case each wave is damped in the direction of propagation.  The
solution contains a primary wave traveling in one direction (first term
on the RHS), whose damping produces a secondary wave traveling in the
opposite direction (second term on the RHS). Notice that when two
primary waves are present ($a_\pm\ne 0$), the profile of ${\bf u}_n$ and
${\bf b}$ will be determined by their interference in a non-trivial
way.

\subsection{Wave reflection by density gradients}
\label{reflection}
We then considered a non-uniform medium where density gradients imply
changes in the Alfv\'en speed, but we neglected ambipolar diffusion.
Eq.~(\ref{evol}) becomes
\be
\label{evolzpgradens}
\pm y_\pm^\prime=iW\tilde\rho^{1/2}y_\pm
\mp\frac{\tilde\rho^\prime}{4\tilde\rho}(y_\pm-y_\mp).
\ee
In terms of the scaled amplitude $f_\pm\equiv \tilde\rho^{1/4}y_\pm$
and the scaled wavevector $q\equiv \tilde\rho^{1/2}W$,
Eq.~(\ref{evolzpgradens}) becomes
\be
f_\pm^\prime=\pm iq f_\pm +\frac{q^\prime}{2q}f_\mp.
\label{v93}
\ee
The general properties of this equation have been discussed in detail  
by Velli~(1993). Here, it is sufficient to notice that for a weak 
density gradient a Taylor expansion gives $q(\xi)\approx W+q_1 \xi$,
which, inserted into Eq.~(\ref{v93}), gives the approximate solution
\be
f_{\pm}\approx (a_{\pm} \mp i\beta a_{\mp})e^{\pm iW\xi} 
\pm i\beta a_{\mp}e^{\mp iW\xi},
\label{approx}
\ee
where $\beta=q_1/4W^2$ and $a_{\pm}$ are constants. Thus, a density
gradient produces two effects: a variation of the wave amplitude, as
expressed by the $\rho^{1/4}$ scaling (Wal\'en 1944) valid in the WKB
approximation, and the generation of secondary reflected waves (second
term on RHS). Notice that this approximate solution is similar (waves traveling in opposite directions) to
Eq.~(\ref{etasol}), thus, again, when two primary waves are present
($a_\pm\ne 0$) the general solution for ${\bf u}_n$ and ${\bf b}$ will
depend on their interference in a non-trivial way.

In the absence of damping, the wave amplitude variation is governed by
the conservation of the energy flux $S=S_+-S_-=\mathrm{const}$, where
$S_{\pm}=|f_\pm|^2/8$ are the fluxes carried by the individual
modes. For a weak gradient, the amplitudes corresponding to the
solutions~(\ref{approx}) are
\be 
|f_{\pm}|^2 = a_{\pm}^2+4\beta a_+a_-\sin(W\xi)\cos(W\xi).
\ee
Thus the density gradient generates reflected waves that induce spatial
oscillations in the wave amplitude, with half the wavelength of the
incident wave, but of course at the same time conserves the energy flux
($S=(a_+^2-a_-^2)/8$). For a given wavelength, the stronger the
gradients, the stronger the coupling, and the reflected wave tends to
have the same amplitude as the incident wave. For a weak gradient the
coupling is negligible and the solution approaches the WKB solution.

\section{Application to molecular cloud cores}
\label{cloud}
\begin{figure}[t]
\centering
\includegraphics[width=\columnwidth]{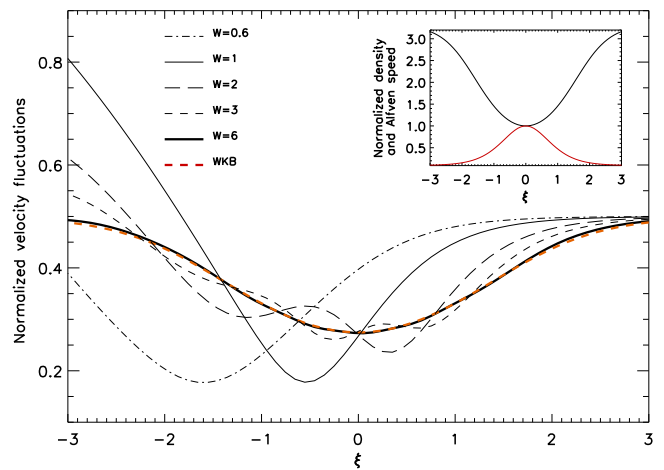}
\caption{Large panel: normalized amplitudes of velocity fluctuations as a function of
$\xi=x/L$ for waves incident on the right side of a core with the density
profile given by Eq.~(\ref{rho_core}) for various values of the nondimensional wave
frequency $W$. No dissipation is included.
The {\it dashed red line} shows the WKB solution. Small panel: the {\it red
line} shows the normalized density profile, and the {\it black
line} the normalized Alfv\'en speed .  \label{urigallWtgh}}
\end{figure}

We applied our numerical schemes to the study of the propagation and
damping of Alfv\'en waves in molecular cloud cores. We considered a
one-dimensional cloud consisting of a low-mass core embedded in a
uniform density envelope.  For the cloud's density we assumed a modified
isothermal profile,
\be
\rho=\mu m_H[n_{\rm env}+(n_{\rm core}-n_{\rm env})\;{\rm sech}^2(\xi)],
\label{rho_core}
\ee 
with $\mu=2.36$, $n_{\rm env}=10^4$~cm$^{-3}$, $n_{\rm
core}=10^5$~cm$^{-3}$, and $\xi=x/L$ where $L=0.04$~pc. The slab is threaded by a perpendicular, straight and uniform magnetic field of strength 
70~$\mu$G and we considered the low-ionization regime typical of
molecular cloud, $\rho\approx \rho_n$. 
We assumed a simple ionization law,
\be
\rho_i=C\rho^\kappa,
\ee
with $\kappa=1/2$ and $C=4.5\times 10^{-17}$~cm$^{-3/2}$~g$^{1/2}$.
With these values, and
$\gamma_{in}=2.38\times10^{14}$~cm$^3$~g$^{-1}$~s$^{-1}$ (Pinto \&
Galli~2008), the ambipolar diffusion resistivity is 
\be
\eta_{\rm AD}=3\times 10^{20}\left(\frac{B}{\mbox{100~$\mu$G}}\right)^2
\left(\frac{n}{\mbox{$10^5$~cm$^{-3}$}}\right)^{-3/2}~\mbox{cm$^2$~s$^{-1}$}.
\ee
The effect of wave reflection is illustrated in the following example, 
which describes the dependence of the velocity fluctuations 
on the normalized wave frequency $W$. We neglected dissipation (assuming
$\eta_{\rm AD}=0$ and $\ell_\bot=\infty$) and assumed that only one
outward propagating wave ($z_+$) is injected at the right boundary.  
Solutions are obtained integrating backward from
$\xi=3$ and do not depend on the phase of the injected wave.  The resulting
velocity amplitudes are shown in Fig.~\ref{urigallWtgh} (large panel) for different
values of the nondimensional frequency $W\in[0.6,6]$.  Clearly, for
increasing frequency the solution approaches the WKB analytical
solution and the highest frequency case $W=6$ is indistinguishable from
it. The amplitude profiles show variations on a scale $1/k=V_A/\omega$
that changes according to the Alfv\'en speed profile (shown in the small panel of Fig.~\ref{urigallWtgh}), being smaller in
the denser region $|\xi|\lesssim 1$. This means that small scales are naturally formed in
the core region.\\
The inclusion of diffusive effects requires a careful treatment of the
boundary conditions. The choice of a symmetric density profile, with
vanishing derivative at the cloud's center and at infinity, allows one to
match the asymptotic solutions with linear combinations of the
solutions obtained by retaining only the terms that involve ambipolar
diffusion (see sect.~\ref{SectAD}), which we require to be exponentially
damped waves.  In practice, we assumed that at $x=\infty$ ($x=-\infty$)
only leftward (rightward) propagating waves are present.  The
integration was carried out only in one half of the domain and we
 looked for solutions symmetric with respect to the core's midplane.
Since the gradients vanish at $x=0$, the coupling given by reflection is
absent and we can impose as boundary condition $|\bf {z}_+(0)|=|\bf {z}_-(0)|$
there. We also averaged over eight phase differences
($\Delta\phi=n\pi/4$ with $n=0,1,..,7$) between $\bf {z}_+(0)$ and $\bf {z}_-(0)$
 to mimic the effect of incoherent wave sources at infinity.
\begin{figure}[b]
\centering
\includegraphics[width=\columnwidth]{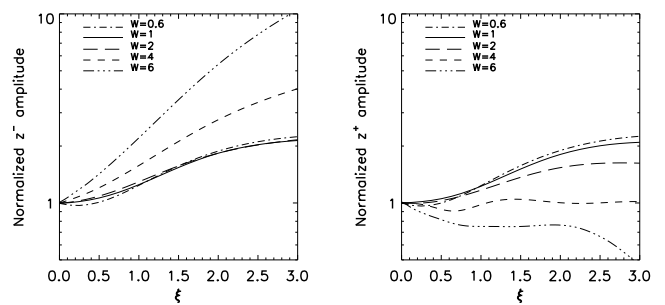}
\caption{Normalized amplitudes of $z_-$ ({\it left panel}\,) and $z_+$
({\it right panel}\,) as a function of $\xi=x/L$ in a cloud core with the density
profile given by Eq.~(\ref{rho_core}) for values of the nondimensional
wave frequency from $W=0.6$ to $W=6$.}
\label{y-allW}
\end{figure}
Results are shown in Fig.~\ref{y-allW} for the range of normalized
frequencies $0.6<W<6$, and over a spatial interval corresponding to
roughly three times the core size. The amplitudes of the waves were normalized to the values in $\xi=0$.  The frequency interval
explored corresponds approximately to the wavelength interval $L \leq
\lambda \leq 10 L$ in the core. In the high-frequency regime ($W=6$),
the damping length is comparable to the scale of the density gradient
($x_d=1.4$ at $\xi=0$), and the dominant effect is the quasi-exponential
damping of the waves on a scale $\sim x_d$.  On the other hand, in the
low-frequency regime ($W=0.6$) the evolution of the wave amplitude is
dominated by the the density gradient on a scale $\xi\sim 1$ , because
the damping length is much longer than the core's size ($x_d=140$ at
$\xi=0$).

\begin{figure}[t]
\centering
\includegraphics[width=\columnwidth]{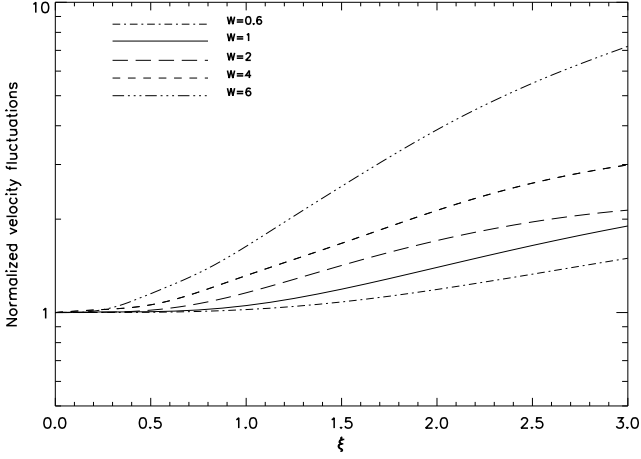}
\caption{Normalized amplitude of
velocity fluctuations as a function of $\xi=x/L$ in a cloud core with the density
profile given by Eq.~(\ref{rho_core}) for values of the nondimensional
wave frequency from $W=0.6$ to $W=6$. }
\label{vphaseW1}
\linespread{1.0}
\end{figure}

Fig.~\ref{vphaseW1} shows the corresponding profiles of velocity
fluctuations for the same range of frequencies as in
Fig.~\ref{y-allW}.  Again an average procedure was performed to take into account relative phases of uncorrelated waves moving in
opposite directions from distant sources. As in the case of the
amplitude of the Els\"asser variables, the velocity amplitudes are
mostly sensitive to the density gradient at lower frequencies $W\sim
0.6$ and mostly sensitive to the ambipolar damping at the highest
frequencies ($W\sim 6$). This is easy to understand given the
dependence of the ambipolar damping length on the inverse square of the
frequency (Eq.~\ref{damp}). Despite the simplicity and idealization of
the model, it is clear that the sharp gradients of velocity dispersion
observed in the environments of dense cores (Barranco \& Goodman~1998;
Caselli et al.~2002; Pineda et al.~2010) cannot be produced by a
density gradient alone, but require the combined effects of a density
gradient and ambipolar diffusion.  In particular, sharp gradients in
velocity amplitude are only possible when $x_d\approx 1$, which implies waves with a frequency such as to give $\ell_d\approx L$.
These waves have wavelength $\lambda \sim (\lambda_{\rm cr} L)^{1/2}$,
intermediate between the critical wavelength $\lambda_{\rm cr}$ and the
typical scale of the density gradient $L$.

\subsection {Heating rate}
\label{heat}
The energy generated by the friction between streaming particles is a
significant source of heating for molecular clouds (Scalo~1977; Lizano
\& Shu~1987). The energy produced heats the bulk of the gas, increasing
its pressure and allowing chemical reactions to proceed at a faster
rate (Flower, Pineau des For\^ets \& Hartquist~1985; Pineau des
For\^ets et al.~1986). Therefore, it is important to evaluate the
heating rate associated to the dissipation of the waves considered in
the previous sections.

In a weakly ionized medium, the rate of energy generation by ambipolar
diffusion (per unit time and unit volume) is
\be
H=\frac{|{\bf B}\times (\nabla\times {\bf B})|^2}
{16\pi^2\gamma_{in}\rho_i\rho_n}
\label{hrate1}
\ee
(Braginskii~1965, see also Pinto et al.~2008).  In the special case of
low-amplitude Alfv\'en waves propagating in a one-dimensional medium,
Eq.~(\ref{hrate1}) reduces to
\be
\langle H\rangle =\frac{\eta_{\rm AD}}{8\pi}|{\bf b}^\prime|^2,
\label{hrate2}
\ee
where the brackets indicate the average over time and the prime a space
derivative. 

\begin{figure}[t]
\centering
\includegraphics[width=\columnwidth]{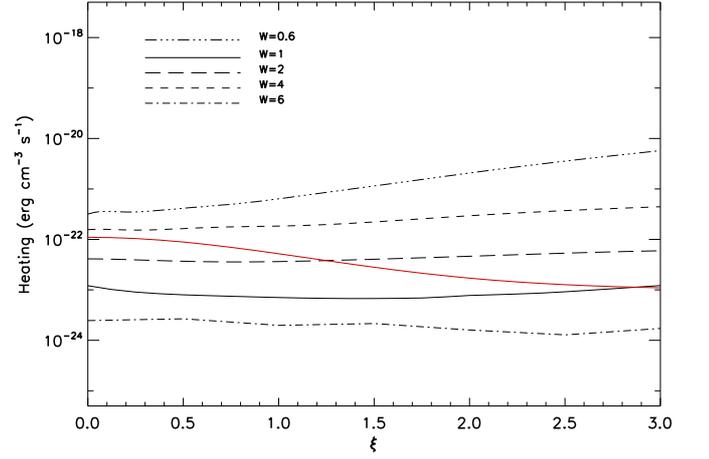}
\caption{Ambipolar diffusion heating as a function of the normalized distance from
the core center. For comparison, the {\em red line} shows the cosmic-ray
heating rate for a cosmic-ray ionization rate $3\times 10^{-17}$~s$^{-1}$
and energy deposition $\Delta Q=20$~eV. }
\label{heatingallW}
\linespread{1.0}
\end{figure}

The heating rate as function of position for waves propagating in the
density profile given by Eq.~(\ref{rho_core}) is shown in
Fig.~\ref{heatingallW}, as before for waves in the frequency range $0.6
\leq W \leq 6$.  To show the results in physical units, we have assumed
$b=B$.  Therefore, the values shown in the figure should be considered as
upper limits to the actual value of the heating rate. For comparison,
the red curve shows the cosmic-ray heating rate computed assuming a
cosmic-ray ionization rate $3\times 10^{-17}$~s$^{-1}$ and a mean
energy deposited per ionization $\Delta Q= 20$~eV (Goldsmith~2001). The
ambipolar diffusion heating is clearly important both in the core and
in the envelope, and competes with cosmic rays as a heating source for
the cloud, especially for the highest values of the frequency. The
behavior of the heating rate deviates significantly from the profile
obtained assuming $b^\prime \sim kb$ (see, e.g., Hennebelle \&
Inutsuka~2006), because of the modulation of the wave amplitude by the
density gradient, and the space dependence of the damping length.  

\subsection{Energy dissipation by turbulent cascade}
In the previous section we have considered the dissipation of linear
Alfv\'en waves caused by ion-neutral collisions.  Here we estimate the
effect of turbulent dissipation as given by the phenomenological
expression~(\ref{phen}). We decided to switch off ambipolar diffusion to better explore the effect 
of turbulence, varying its strength through the control parameter $\tau_A/\tau_{NL}$. Setting
$\eta_{AD}=0$ we set $\lambda_{cr}=0$ and we did not have to worry about 
the first inequality in Eq.~(\ref{hp}). 
We imposed at both sides of the cloud incoming fluctuations
characterized by a frequency spectrum of width $1/t_A$ 
and then falling as  $\omega^{-2}$: energy is at long wavelengths and
$\ell_{||}\approx V_{A0} t_A= L$. 
We here discuss four runs in which we varied the turbulent strength
$\tau_A/\tau_{NL}$ by changing the input amplitude and
perpendicular scale as listed in Table~\ref{table1}. 
The strong turbulence cases are runs A and D, the last even though it has a smaller
input amplitude. The time-averaged, turbulent heating rate per unit volume follows 
directly from Eq.~(\ref{phen}) and is given by 
\be
H_T=\rho_n\frac{|z_-||z_+|^2+|z_+||z_-|^2}{2\ell_\bot}
\propto\rho_n f(\sigma_c)\frac{E^{3/2}}{2\ell_\bot},
\label{diss}
\ee
where 
\be
\sigma_c=\frac{|z_+|^2-|z_-|^2}{|z_+|^2+|z_-|^2}
\ee
 is the (nomalized) cross helicity, $E=|z_+|^2+|z_-|^2$ is the 
 total turbulent energy, and $f(\sigma_c)=\sqrt{1-\sigma_c^2}
(\sqrt{1+\sigma_c}+\sqrt{1-\sigma_c})$ is a symmetric function that
vanishes at $|\sigma_c|=1$ and increases monotonically, reaching a maximum
$f(0)=1$.\\
The small amount of energy at high frequency justifies neglecting ambipolar
diffusion, we verified that $\langle H\rangle<<H_T$ for
$\tilde{\eta}_{AD}\simeq\lambda_{cr}/\ell_{||}\lesssim\textrm{min}(b/B)=0.1$.

\begin{figure}[t]
\centering
\includegraphics[width=0.98\linewidth]{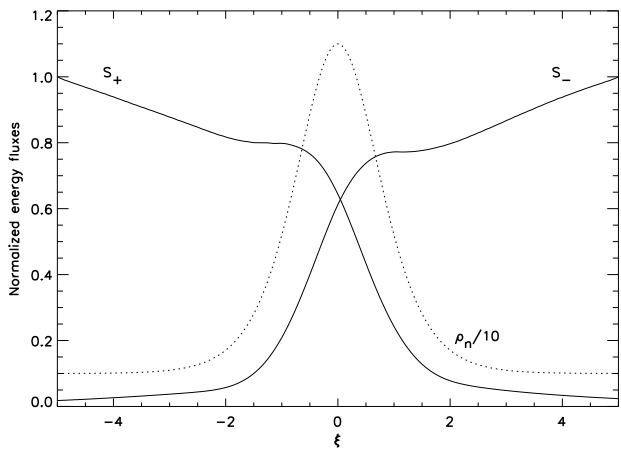}
\caption{Time-averaged normalized energy fluxes entering the cloud from
both sides, $S_+/S_+(\xi=-5)$ and $S_-/S_-(\xi=5)$, for run B. 
The normalized density profile is shown for comparison by the {\it dotted line}.} 
\label{turbflux}
\end{figure}

Let us examine the case of intermediate turbulence strength,
$\tau_A/\tau_{NL}=1/\sqrt{10}$ (run B).
In the linear case with $\eta_{AD}=0$, the coupling between $z_\pm$ in proximity of the density
gradients causes an enhancement of the respective energy fluxes
$S_\pm$, which have a bump toward the cloud's center. This ensures the
conservation of the \emph{net} energy flux ($S_+ - S_-$) which is uniform. When
nonlinear dissipation is taken into account,
the profiles of $S_+$ and $S_-$ are remarkably different, as shown in
Fig.~\ref{turbflux} where we also plot the density profile for comparison
(dotted line). Both energy fluxes
decrease toward the cloud's center, indicating the damping of the wave
(heating). 
In contrast to the linear case, there is a sharp decrease on the
core size due to the turbulent heating $H_T$, which is stronger in the whole
core (being approximately proportional to $\rho_n$). From Fig.~\ref{turbflux}
one can also deduce the behavior of $\sigma_c$, which is also equal to
$(S_+-S_-)/(S_++S_-)$ (the
fluxes are normalized to their respective incident value, since 
$S_+(\xi=-5)\approx S_-(\xi=5)$ the normalization factors can be considered
equal). The normalized cross helicity vanishes at the
core's center and increases rapidly in the core's envelope, maintaing a value
$|\sigma_c|\gtrsim 0.8$ at the edges of the core.
\begin{figure}[t]
\centering
\includegraphics[width=0.98\linewidth]{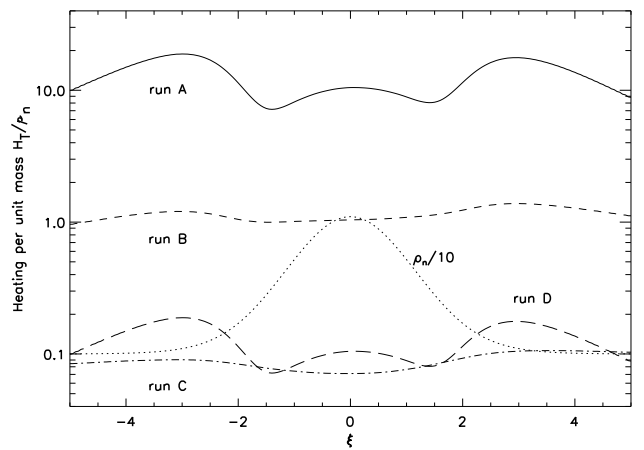}
\caption{Profile of the time-averaged heating rate per unit mass (in nondimensional units) for
different turbulent strength $\tau_A/\tau_{NL}=1,~1/\sqrt{10},~1/10,~1$, corresponding to runs
 A, B, C, D as labeled on each curve ({\it solid, dashed, dot-dashed and long-dashed lines,
respectively}). 
The normalized density profile is shown for comparison as a {\it
dotted line}.}
\label{heating}
\end{figure}
\begin{table}[t]
\begin{center}
\begin{tabular}{cccc}
RUN & $b/B$ & $\ell_\bot/\ell_{||}$ & $\tau_a/\tau_{NL}$ \\
\hline
A   &   1   &   1   &   1  \\
B   & $1/\sqrt{10}$  &   1   & $1/\sqrt{10}$ \\
C   & 1/10  &   1   & 1/10 \\
D   & 1/10  &  1/10 &   1  \\
\hline
\end{tabular}
\end{center}
\caption{Nondimensional parameters for the runs: input amplitude ($b/B$) and
ratio of the perpendicular to parallel scales of fluctuations
($\ell_\bot/\ell_{||}$). The turbulence strength 
$\tau_A/\tau_{NL}=b\ell_{||}/B\ell_\bot$ is
also indicated in the last column.} 
\label{table1}
\end{table}
The cross helicity and the total turbulent energy affect the average heating 
rate per unit mass ($H_T/\rho_n$), which is plotted in Fig.~\ref{heating} 
for the four runs (run B corresponding to the dashed line). For weak turbulence
 (runs B and C) the profile is basically flat, varying by 20\% at maximum,
  and thus reflecting $H_T\propto \rho_n$. Interestingly, it is flat on a region 
  entirely surrounding the core and peaks outside it, showing that the wave 
  damping is maximal when passing from the cloud to the core. 
The relatively flat profile can be explained in terms of a compensation
of the terms related to the cross-helicity and total turbulent energy in
Eq.~(\ref{diss}). 
Indeed, the cascade is more balanced ($\sigma_c\approx 0$) at the
center, but less energetic, while the turbulent energy grows outside the core, 
but the cascade is less balanced.
For stronger turbulence (e.g. $\tau_A/\tau_{NL}=1$) these two effects
do not compensate each other anymore and the minima and
maxima in the heating rate are more pronounced ($H_T$ 
follows the scaling with $\rho_n$ less closely).
Note that keeping $\tau_A/\tau_{NL}=1$ but varying the input amplitudes
$b/B=\ell_\bot/\ell_{||}=1/10$ the profile does not vary, but the overall
dissipation decreases (compare runs A and D). One can verify that the 
average dissipation per unit mass scales as
$H_T/\rho_n\propto1/\tau_{NL}^2$, while  
the relative excursion at the core edges 
$\Delta H_T/H_T$ grows with $\tau_{NL}/\tau_{A}$.

\section{Conclusions}
\label{concl}
We have considered the propagation of Alfv\'en waves in a density
profile typical of a molecular cloud core embedded in a low-density
medium. The formulation of the problem in terms of Els\"asser variables
was found particularly appropriate to treat the effects of wave
reflection and dissipation. We have studied with time-dependent
as well as time-independent numerical codes the interaction of incoherent wave
trains incident on the opposite edges of a one-dimensional model cloud,
determining the amplitude variations associated to the reflection and
damping of the waves.

We found that for waves with a wavelength longer than the critical
wavelength, of the order of a few $10^{-3}$~pc in a molecular cloud
core, but shorter or of the order of the size of the core, a
significant decrease in the velocity amplitude is produced by the
combined effects of reflection at the core's boundaries and dissipation
due to ambipolar diffusion in the core's interior. For these
waves, the damping length associated to ion-neutral collisions is of
the same order as the size of the core. The resulting behavior of the
amplitude of velocity fluctuations may explain, at least in part, the
sharp decrease of line widths observed in the environments of low-mass
cores (see sect.~\ref{intro}). Moreover, for these waves the
energy generated by ion-neutral drift is a significant source of
heating both in the core and its envelope (see sect.~\ref{heat}).

We also considered the turbulent dissipation and found that the average
heating per unit mass scales as $H_T/\rho_n\propto
(b/\ell_\bot)^2=\tau_{NL}^{-2}$, thus becoming significant for high enough wave
amplitudes ($0.1<b/B<1$). This regime is
particularly relevant for molecular clouds and star-forming regions
where a statistical analysis of dust polarization maps suggests that the
ratio of the strength of turbulent magnetic field to the large-scale
mean field is of the order of 0.1--0.5 (Hildebrand et al.~2009; Houde
et al.~2009). We used a phenomenological term to mimic the effect of a
perpendicular turbulent cascade.  The results of our calculations show
that the turbulent cascade affects the amplitude of fluctuations both
at the cloud's edge and in the cloud's interior. In particular, the
energy dissipation (per unit mass) associated to this process is peaked in the
regions immediately surrounding the dense central core, indicating a stronger
damping of the wave amplitude. The relative excursion
between maxima and minima $\Delta H_T/H_t$ depends on the
turbulence strength $\tau_{NL}/\tau_A$,
so it can be significant even for small amplitudes, provided the turbulent
correlation length is of the order of a few times the size of the core.  
In a more realistic situation, the turbulent cascade, although it is stronger in the perpendicular direction, 
also generates small parallel scales. Therefore we expect that turbulence and ambipolar 
diffusion will be at work at the same time, enhancing the sharp variation of velocity
amplitude at the interface between the cloud and the core. In the 
present work the investigation of the parameter space was limited by the several
assumptions (Eq.~(\ref{hp})) we made to obtain the model equations. Interesting
 features appear when we are close to the limit of the allowed parameter space.
Additional work is needed to address these issues with a more realistic and
detailed modeling of turbulence that includes the effects of ambipolar
diffusion.

\acknowledgements
We would like to thank Stella Galligani for carrying out some of the nonlinear simulations with the phenomenological model of turbulence.

\end{document}